\documentclass[aps,prd,showpacs,onecolumn,preprintnumbers,notitlepage,
nofootinbib,amsfont,amssymb,10pt,singlespacing] {revtex4-1}
\usepackage{amsmath,bm}
\usepackage{times}
\usepackage{braket}
\usepackage{color,graphicx}
\usepackage{slashed}
\usepackage{CJK,upgreek,fancyhdr}
\usepackage{hyperref}

\newcommand{\beq}{\begin{eqnarray}}
\newcommand{\eeq}{\end{eqnarray}}
\newcommand{\non}{\nonumber\\ }

\newcommand{\psl}{ P \hspace{-2.8truemm}/ }
\newcommand{\nsl}{ n \hspace{-2.2truemm}/ }
\newcommand{\vsl}{ v \hspace{-2.2truemm}/ }
\newcommand{\epsl}{\epsilon \hspace{-1.8truemm}/\,  }

\def\lsim{ {\ \lower-1.2pt\vbox{\hbox{\rlap{$<$}\lower6pt\vbox{\hbox{$\sim$}
}}}\ } }
\def\gsim{ {\ \lower-1.2pt\vbox{\hbox{\rlap{$>$}\lower6pt\vbox{\hbox{$\sim$}
}}}\ } }

\def \plb{  Phys. Lett. B }

\def \prd{  Phys. Rev. D }

\def \jhep{ J. High Energy Phys.  }

\definecolor{Red}{rgb}{1.,0.,0.}

\definecolor{Blue}{rgb}{0.,0.,1.}

\definecolor{nicered}{rgb}{0.7,0.1,0.2}
\definecolor{nicegreen}{rgb}{0.1,0.4,0.2}

\hypersetup{colorlinks,citecolor=nicegreen,linkcolor=nicered}

\begin{document}
\begin{CJK*}{GB}{gbsn}
\title{\boldmath Improved perturbative QCD formalism for $B_c$ meson decays}
\author{Xin~Liu}
\email
{liuxin@jsnu.edu.cn}
\affiliation{School of Physics and Electronic Engineering,\\
Jiangsu Normal University, Xuzhou 221116, People's Republic of China}

\author{Hsiang-nan~Li}
\email
[Corresponding author: ]
{hnli@phys.sinica.edu.tw }
\affiliation{ Institute of Physics, Academia Sinica, Taipei, Taiwan 115, Republic of China}

\author{Zhen-Jun~Xiao}
\email
{xiaozhenjun@njnu.edu.cn}
\affiliation{ Department of Physics and Institute of Theoretical
Physics,\\
Nanjing Normal University, Nanjing 210023, People's Republic of China}


\date{\today}

\begin{abstract}

We derive the $k_T$ resummation for doubly heavy-flavored $B_c$ meson
decays by including the charm quark mass effect into the
known formula for a heavy-light system. The resultant Sudakov factor
is employed in the perutrbative QCD study of the ``golden channel"
$B_c^+ \to J/\psi \pi^+$. With a reasonable model for the $B_c$ meson
distribution amplitude, which maintains approximate on-shell conditions
of both the partonic bottom and charm quarks, it is observed that the
imaginary piece of the $B_c \to J/\psi$ transition form factor appears
to be power suppressed, and the $B_c^+ \to J/\psi \pi^+$ branching ratio
is not lower than $10^{-3}$. The above improved perutrbative QCD formalism
is applicable to $B_c$ meson decays to other charmonia and charmed mesons.

\end{abstract}


\maketitle

\section{INTRODUCTION}

\begin{figure}[!!htb]
\centering
\begin{tabular}{l}
\includegraphics[width=0.7\textwidth]{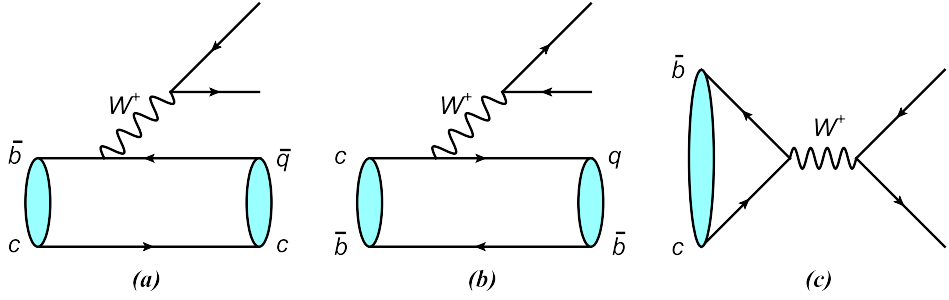}
\end{tabular}
\caption{(Color online) Diagrams for $B_c$ meson decays. }
\label{fig:fig1}
\end{figure}

A $B_c$ meson is the ground state of the doubly heavy-flavored $\bar b c$
system in the Standard Model~\cite{Brambilla:2004wf}, different from
the heavy-light one represented by a $B$ meson and from the heavy-heavy
one represented by quarkonia $J/\psi$ and $\Upsilon$
in many aspects. Its weak transition can occur through the bottom quark
decay with the spectator charm quark as displayed in Fig.~\ref{fig:fig1}(a), the
charm quark decay with the spectator bottom quark in Fig.~\ref{fig:fig1}(b), and
the pure weak annihilation channel in Fig.~\ref{fig:fig1}(c).
Hence, $B_c$ meson decays contain rich heavy quark dynamics in both the perturbative
and nonperturbative regimes, which is worth a thorough exploration
with high precision. It is certainly a challenge to develop an appropriate
theoretical framework for analyzing $B_c$ meson decays. A framework
available in the literature is the perturbative QCD (PQCD) approach, which
basically follows the conventional one for $B$ meson decays,
with the finite charm quark mass being included in hard decay kernels
but neglected in the $k_T$ resummation for meson distribution amplitudes.
A rigorous resummation formalism for $B_c$ meson decays, which involve
multiple scales, is expected to be more complicated than for $B$ meson decays.

In this paper, we will investigate how the charm quark mass affects the infrared
structures of the $B_c$ meson and of its decay products and derive the
corresponding $k_T$ resummation in the PQCD approach. The
derivation depends on the power counting for the ratio $m_c/m_{b}$, $m_b$ ($m_c$)
being the bottom (charm) quark mass. Taking the limit $m_b\to\infty$ but
keeping $m_c$ finite, we treat a $B_c$ meson as a heavy-light system,
the decays of which can be analyzed in the conventional PQCD approach to $B$ meson
decays mentioned above. Taking the limit $m_b,m_c\to\infty$ but fixing the
ratio $m_c/m_b$, we treat a $B_c$ meson as a heavy-heavy system, the
decays of which may be studied in a formalism for heavy quarkonium
decays. Here, we will adopt the power counting rules proposed
in Ref.~\cite{Kurimoto:2002sb} and regard a $B_c$ meson as a multiscale system,
which respects the hierarchy $m_b\gg m_c\gg \Lambda_{\rm QCD}$,
$\Lambda_{\rm QCD}$ being the QCD scale. An intermediate impact of this power
counting is that the large infrared logarithms $\ln(m_b/m_c)$, in addition
to the ordinary ones $\ln(m_b/\Lambda_{\rm QCD})$, appear in the perturbative
evaluation of the $B_c$ meson distribution amplitude and need to be resummed.

The Sudakov factor from the $k_T$ resummation with the charm quark mass effect
is then employed in the PQCD study of the ``golden channel"
$B_c^+ \to J/\psi \pi^+$. We focus on the bottom quark decay of a $B_c$ meson
because the charm quark decay is believed to suffer from significant
long-distance contributions, i.e., final-state interactions, though perturbative
results for the $B_c \to B_{(s)} X$ modes have been presented in the
literature~\cite{Sun:2014ika,Sun:2015exa}. Besides, the available models for the
$B_c$ meson distribution amplitude vary dramatically from a simple
$\delta$ function~\cite{Cheng:2005um,Bell:2008er} to a complicated Gaussian
type~\cite{Sun:2014ika}. We will propose a kinematic constraint
on the $B_c$ meson distribution amplitude, which allows both the partonic bottom
and charm quarks to be off shell only at a power-suppressed level. It is then
shown, with a reasonable model for the $B_c$ meson distribution amplitude,
that the imaginary piece of the $B_c \to J/\psi$ transition form factor, supposed
to be a real object~\cite{Manohar:2000dt}, is indeed power suppressed. It is also
found that the $B_c^+ \to J/\psi \pi^+$ branching ratio is not lower than $10^{-3}$,
in agreement with those obtained in other approaches.

In Sec.~II, we discuss the kinematic constraint on the charm quark momentum
distribution in a $B_c$ meson. The one-loop correction to the $B_c$ meson
distribution amplitude, which generates the double logarithm
$\alpha_s\ln^2(m_b/m_c)$, $\alpha_s$ being the strong coupling, is
calculated in Sec.~III. The result hints at how the $k_T$ resummation for
$B_c$ meson decays is modified from the known formula for $B$ meson decays.
In Sec.~IV, we predict the $B_c^+ \to J/\psi \pi^+$ branching ratio in the
improved PQCD framework, including the contributions from both factorizable
and nonfactorizable emission diagrams. It is then stressed in the Conclusion
that the formalism developed here is ready for the extension to $B_c$ meson
decays to other charmonia like $\eta_c$, $\chi_{cJ}$ $(J=0,1,2)$, \ldots, and
charmed mesons.

\section{\boldmath Kinematic constraint on $B_c$ meson distribution amplitude}

Consider the $B_c(P_1)\to J/\psi(P_2)$ transition at the maximal recoil,
where
\begin{eqnarray}
P_1 = \frac{m_{B_c}}{\sqrt{2}} (1, 1, {\bf 0}_T)\;, \qquad
P_2 = \frac{m_{B_c}}{\sqrt{2}} (1, r_{J/\psi}^2, {\bf 0}_T) \label{bj}
\end{eqnarray}
in the light-cone coordinates label the $B_c$ and $J/\psi$ meson momenta,
respectively, with $r_{J/\psi} = m_{J/\psi}/m_{B_c}$ and
$m_{B_c}$ ($m_{J/\psi}$) being the $B_c$ ($J/\psi$) meson mass.
This transition involves multiple scales the same as in the $B\to D^*$
transition, which has been studied in Ref.~\cite{Kurimoto:2002sb}: $m_b$ from
the initial-state $B$ meson, $m_c$ from the final-state $D^*$ meson, and
both the $B$ and $D^*$ bound states contain the nonperturbative
dynamics characterized by a low hadronic scale $\Lambda$. Following the
argument in Ref.~\cite{Kurimoto:2002sb}, the scaling of the energetic $J/\psi$ momentum
$P_2\sim (m_b, m_c^2/m_b,{\bf 0}_T)\sim m_c(m_b/m_c, m_c/m_b,{\bf 0}_T)$
hints that the components of a collinear gluon momentum in such a multiscale
system also obeys the power counting
\begin{eqnarray}
l^\mu\sim \left(\frac{m_b}{m_c}\Lambda,\frac{m_c}{m_b}\Lambda,
\Lambda\right),
\end{eqnarray}
with a tiny invariant mass squared $l^2\sim {\cal O}(\Lambda^2)$.
A valence charm quark in the $J/\psi$ meson, after emitting such a collinear
gluon, can acquire the virtuality of order $P_2\cdot l\sim m_c\Lambda$.
The momentum parametrizations for the two valence charm quarks
participating in the hard subprocess
should be symmetric under their exchange. Denote the spectator charm
quark momentum as $k_2=x_2P_2$ and another as $P_2-k_2=(1-x_2)P_2$ with the
momentum fraction $x_2$, and assume both of them to be off-shell at most by
${\cal O}(m_c\Lambda)$:
$k_2^2-m_c^2={\cal O}(m_c\Lambda)$ and $(P_2-k_2)^2-m_c^2={\cal O}(m_c\Lambda)$.
To satisfy these two conditions simultaneously, we choose a charm quark
mass $m_c\approx m_{J/\psi}/2\sim 1.5$ GeV for $m_{J/\psi}= 3.097$ GeV,
and the momentum fraction $x_2=1/2\pm\delta$ can deviate from its central
value by $\delta\sim {\cal O}(\Lambda/m_c)$. That is,
the $J/\psi$ distribution amplitude takes a substantial value in the
above range of $x_2$ with $\delta\sim 0.3$ for $\Lambda\sim 0.5$ GeV,
due to the effect of collinear gluon emissions. The model for the
$J/\psi$ meson distribution amplitude, proposed in Ref.~\cite{Bondar:2004sv} and
widely employed in the PQCD analyses, does exhibit these features.

Next, we discuss the kinematic constraint on the shape of the $B_c$ meson
distribution amplitude. Label the momentum of the spectator charm quark
in the $B_c$ meson by $k_1$ and that of the bottom quark by $P_1-k_1$.
The approximate on-shell-ness of the partons, $k_1^2\sim m_c^2$
and $(P_1-k_1)^2\sim m_b^2$, implies that the zeroth component
of $k_1$ is of order $k_1^0\sim m_c$. A $B_c$ meson at rest is dominated
by soft dynamics, for which the momentum of a soft gluon is characterized
by the power counting~\cite{Kurimoto:2002sb}
\begin{eqnarray}
l^\mu\sim \left(\Lambda,\Lambda,\Lambda\right),
\end{eqnarray}
with a tiny invariant mass squared $l^2\sim {\cal O}(\Lambda^2)$.
The spectator charm quark, after emitting such a soft gluon,
then reaches the virtuality of order $k_1\cdot l\sim m_c\Lambda$.
Parametrize the charm quark momentum by $k_1=x_1P_1$,
$x_1$ being a momentum fraction, and require the virtuality
$k_1^2-m_c^2={\cal O}(m_c\Lambda)$. Given the bottom quark mass
$m_b\approx m_{B_c}-m_c\sim 4.8$ GeV for $m_{B_c}=6.276$ GeV,
we find that the $B_c$ meson distribution amplitude takes a substantial
value around the momentum fraction $x_1\sim m_c/m_b\sim 0.3$
within the width of about $\Lambda/m_b\sim 0.1$. It can be verified,
following the above discussion, that the bottom quark in the $B_c$
meson acquires the virtuality of $(P_1-k_1)^2-m_b^2\sim {\cal O}(m_b\Lambda)$,
consistent with the soft gluon emission effect.

We then investigate the virtuality of the hard particles in the kinematic
regions specified for the partonic bottom and charm quarks. First, the
invariant mass of the hard gluon emitted by the spectator quark is written as
\begin{eqnarray}
(k_1-k_2)^2\approx-\frac{m_bm_c}{2}+{\cal O}(m_b\Lambda),
\label{gluon}
\end{eqnarray}
with the insertion of $m_{B_c}\approx m_b+m_c$ and $m_{J/\psi}\approx 2m_c$
up to the first powers in $m_c$ and in $\Lambda$. The first term on the right-hand side of
Eq.~(\ref{gluon}), being ${\cal O}(m_bm_c)$, indicates that the hard gluon
tends to be spacelike for the chosen mass scales $m_b$, $m_c$, and $\Lambda$.
The hard bottom quark, to which the hard gluon attaches, remains spacelike
with the virtuality
\begin{eqnarray}
(P_1-k_2)^2-m_b^2\approx -\frac{m_b^2}{2}.
\end{eqnarray}
The hard charm quark, to which the hard gluon attaches, is also
spacelike with the virtuality
\begin{eqnarray}
(P_2-k_1)^2-m_c^2\approx -m_bm_c+{\cal O}(m_b\Lambda).
\end{eqnarray}
We conclude that, as both the partonic
bottom and charm quarks are only off-shell a bit, the imaginary piece in the
$B_c \to J/\psi$ transition form factor appears to be power suppressed.
This observation is easily understood: the $J/\psi$ meson mass
is below the $D\bar D$ threshold, so the $B_c \to J/\psi$ transition
hardly occurs through an intermediate state.

\begin{figure}[!!htb]
\centering
\begin{tabular}{l}
\includegraphics[width=0.53\textwidth]{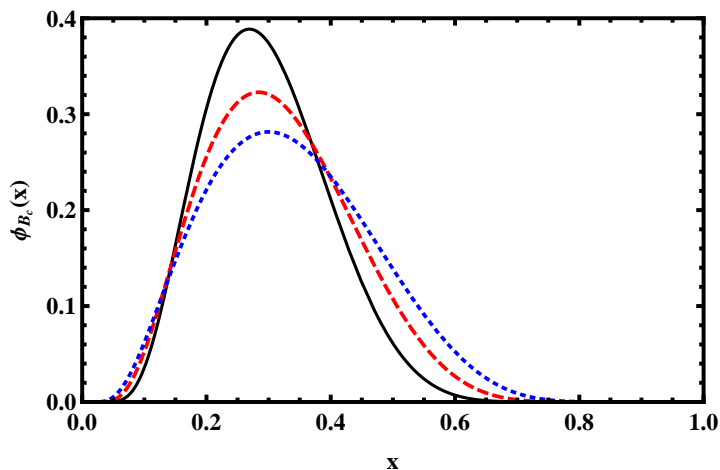}
\end{tabular}
\caption{(Color online)
Behavior of $\phi_{B_c}(x)$ for the different shape parameters
$\beta_{B_c} = 0.8$ GeV (black-solid curve), 1.0 GeV (red-dashed curve),
and 1.2 GeV (blue-dotted curve).}
\label{fig:fig2}
\end{figure}

The $B_c$ meson wave function with an intrinsic $k_T$
dependence is parametrized in a Gaussian form as~\cite{Huang80}
\beq
\phi_{B_c}(x,k_T) &=& \frac{f_{B_c}}{2\sqrt{2 N_c}}
\frac{\pi}{2\beta_{B_c}^2}N_{B_c}\exp\left[-\frac{1}{8\beta_{B_c}^2}
\left(\frac{|{\bf k}_T|^2 + m_c^2}{x}+ \frac{|-{\bf k}_T|^2+m_b^2}{1-x}\right)\right]\;,
\eeq
in which ${\bf k}_T$ $(-{\bf k}_T)$ is the transverse momentum carried by
the charm (bottom) quark, $N_c$ is the number of colors, $\beta_{B_c}$ is
the shape parameter, and $N_{B_c}$ is the normalization constant. The $B_c$
meson distribution amplitude is given by
\begin{eqnarray}
\phi_{B_c}(x,b)&=&  \frac{f_{B_c}}{2\sqrt{2 N_c } }
N_{B_c}x(1-x)\exp\left[-\frac{(1-x)m_c^2+xm_b^2}
{8\beta_{B_c}^2x(1-x)}\right]\exp\left[-2\beta_{B_c}^2x(1-x)b^2\right],
\end{eqnarray}
with the impact parameter $b$ being conjugate to $k_T$.
The normalization constant $N_{B_c}$ is fixed by the
relation
\beq
\int_0^1 \phi_{B_c}(x,b=0) dx \equiv\int_0^1 \phi_{B_c}(x) dx= \frac{f_{B_c}}{2\sqrt{2 N_c} },
\eeq
where the decay constant $f_{B_c} = 0.489 \pm 0.005$ GeV has been obtained
in lattice QCD by the TWQCD Collaboration~\cite{Chiu:2007km}.
Figure~\ref{fig:fig2}, in which the behavior of $\phi_{B_c}(x)$
is plotted for the different shape parameters $\beta_{B_c}$, indicates
that the peak of $\phi_{B_c}(x)$ shifts toward larger $x$ and becomes
broader with the increase of $\beta_{B_c}$. Note that data for $B_c$ meson
decay branching ratios are not yet available, so it is difficult to
determine $\beta_{B_c}$ unambiguously.
However, the kinematic constraint derived above hints that
$\beta_{B_c}=1.0$ GeV seems to be a reasonable choice.
On the other hand, the existent
models~\cite{Sun:2014ika,Wang:2017bgv} of the $B_c$ meson distribution
amplitude roughly correspond to the range [0.6, 1.0] GeV of the
parameter $\beta_{B_c}$.

\section{\boldmath $k_T$ resummation for $B_c$ meson decays}

\begin{figure}[!!htb]
\centering
\begin{tabular}{l}
\includegraphics[width=0.8\textwidth]{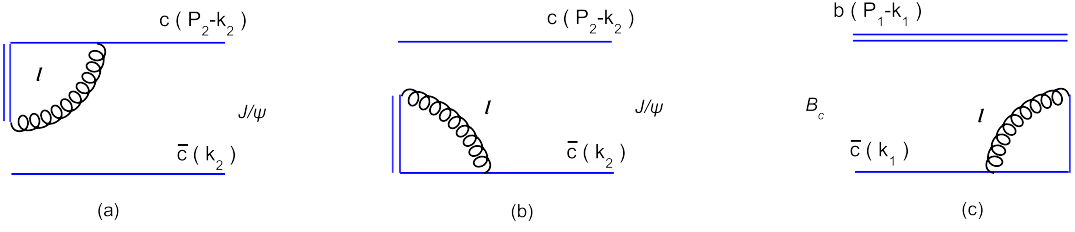}
\end{tabular}
\caption{(Color online) ${\cal O}(\alpha_s)$ effective
diagrams for the $J/\psi$ and $B_c$ mesons wave functions,
which are relevant to the Sudakov factor $s_c(Q,b)$.}
\label{fig:fig3}
\end{figure}

A theoretical challenge from the $B_c\to J/\psi$ transition
is to derive the $k_T$ resummation for energetic charm quarks
with a finite mass. To proceed, we construct a
transverse momentum-dependent $J/\psi$ meson wave function in
the $k_T$ factorization theorem~\cite{Nagashima:2002ia,Li:2004ja}
and then perform the perturbative evaluation according to the wave-function
definition as a hadronic matrix element of a nonlocal operator.
The double logarithms attributed to the overlap of the collinear and
soft radiative corrections are expected to differ from those in $B$ meson
decays into light mesons, which have been elaborated in Ref.~\cite{Li:1994cka}.
According to the one-loop analysis in Ref.~\cite{Li:2000hh}, the only source
of the double logarithms is the correction to the quark-Wilson-line
vertex as displayed in Fig.~\ref{fig:fig3}(a), in which the loop momentum
does not flow into a hard subprocess. When the gluon in Fig.~\ref{fig:fig3}(a)
attaches to the lower piece of the Wilson lines, the loop momentum flows
through a hard subprocess. Since the region with small parton momenta
dominates in the $k_T$ factorization, the large collinear gluon momentum
induces power suppression on the hard kernel~\cite{Li:2004ja}, such that this
one-loop diagram does not generate the double logarithm. The similar vertex
diagram with the gluon being radiated by the spectator charm quark
either in the $J/\psi$ meson [Fig.~\ref{fig:fig3}(b)] or in the $B_c$ meson
[Fig.~\ref{fig:fig3}(c)] may produce the double logarithms. Nevertheless, their
effects ought to be weaker, due to the lack of phase space for collinear
gluons from less energetic quarks.

The loop integral corresponding to Fig.~\ref{fig:fig3}(a) is written as
\begin{eqnarray}
\phi^{(1)}=-\frac{i}{4}g^2C_F\mu_{\rm
f}^{2\epsilon} \int\frac{d^{4-2\epsilon}l}{(2\pi)^{4-2\epsilon}}
tr\left[ \gamma_5\not n_+\frac{\not \bar k+\not l+m_c}{(\bar
k+l)^2-m_c^2}\gamma_\nu\not n_-\gamma_5\right]
\frac{1}{l^2}\frac{n^\nu}{n\cdot l}, \label{pkd}
\end{eqnarray}
with $\bar k\equiv P_2-k_2$, the eikonal vertex $n_\nu$, and the eikonal
propagator $1/n\cdot l$. The dimensionless vector $n$ with $n^+ > 0$
represents the direction of the Wilson lines, which is allowed to be away
from the light cone~\cite{Li:2000hh}. The projectors $\gamma_5 \nsl_+$
and $\nsl_-\gamma_5$, arising from the insertion of the Fierz identity
for factorizing the fermion flow, work for the selection of the logarithm
$\ln (m_b/m_c)$ up to corrections in powers of $m_c/m_b$. A straightforward
calculation leads to
\begin{eqnarray}
\phi^{(1)}=\frac{\alpha_s}{4\pi} C_F \biggl[ \frac{1}{\epsilon} +
\ln{\frac{4\pi \mu_f^2}{m_c^2 e^{\gamma_E}}} -\ln^2\frac{\zeta^2}{k_T^2}
+\ln^2\frac{m_c^2}{k_T^2}+\ln\frac{\zeta^2}{m_c^2} +2- \frac{2\pi^2}{3}\biggr],
\label{mass}
\end{eqnarray}
with the factorization scale $\mu_f$, the Euler constant $\gamma_E$,
and the variable $\zeta^2 \equiv 4(n\cdot \bar{k})^2/n^2$. It is found
that the infrared logarithms in the above expression reproduce those
in the pion case~\cite{Nandi:2007qx}, as $m_c$ is replaced by $k_T$.
The double logarithms can be understood in the way that the soft
divergence is regularized by the quark virtuality $k_T$, and the collinear
divergence is regularized by the charm quark mass $m_c$, giving
\begin{eqnarray}
-\ln^2\frac{\zeta^2}{k_T^2}+\ln^2\frac{m_c^2}{k_T^2}
=-\ln\frac{\zeta^2m_c^2}{k_T^4}\ln\frac{\zeta^2}{m_c^2}.
\end{eqnarray}
The partial cancellation between the two double
logarithms implies that the resummation effect in
the case of energetic massive quarks is smaller than in the case of
light quarks~\cite{Aglietti:2007bp}.

The aforementioned lack of phase space for the collinear gluons in
Figs.~\ref{fig:fig3}(b) and~\ref{fig:fig3}(c) can be
understood by means of the contour integration. Take Fig.~\ref{fig:fig3}(b),
the loop integrand of which contains a denominator $(k_2-l)^2-m_c^2$ from
the anticharm quark propagator, as an example.
To get a nonvanishing contribution from the contour integration over
the minus component $l^-$ of the loop momentum, some poles of $l^-$
have to be located in the upper half-plane, and some have to be located
in the lower half-plane. This is possible only when the coefficients of $l^-$ in the
denominators of the corresponding loop integrand are not of the same sign.
Hence, the plus component $l^+$ must take a value in the range
$0 < l^+ < k_2^+$ for our gauge choice $n^+ > 0$ as stated below Eq.~(\ref{pkd}).
In the dominant region with small parton momenta, i.e., with small $k_2^+$,
the phase space for $l^+$ is then limited, implying a weaker double
logarithmic effect.

We will not attempt a complete one-loop computation and an exact
next-to-leading-logarithm resummation associated with an energetic massive quark in
the present work. Instead, we will infer an approximate Sudakov exponent
from the implication of Eq.~(\ref{mass}). It has been known that the
$k_T$ resummation for an energetic light quark yields the Sudakov exponent
in the $b$ space~\cite{Botts:1989kf,Li:1994cka},
\begin{eqnarray}
s(Q,b)=\int_{1/b}^Q\frac{d \mu}{\mu}
\left[\int_{1/b}^{\mu}\frac{d\bar\mu}{\bar\mu}A(\alpha_s(\bar\mu))
+B(\alpha_s(\mu))\right],\label{light}
\end{eqnarray}
at the next-to-leading-logarithm accuracy, where the universal anomalous
dimension $A(\alpha_s)$ given to two loops is responsible for the collection of the
double logarithms, the factor $B(\alpha_s)$ given to one loop is for the collection
of single logarithms, and $Q$ is related to the
major light-cone component of the quark momentum through the variable $\zeta$.
The $\mu_f$-independent logarithms in Eq.~(\ref{mass}) can be cast into
two pieces,
\begin{eqnarray}
-\left(\ln^2\frac{\zeta^2}{k_T^2}- \ln\frac{\zeta^2}{k_T^2} \right)
+\left(\ln^2\frac{m_c^2}{k_T^2}-\ln\frac{m_c^2}{k_T^2}\right),\label{mass1}
\end{eqnarray}
which are of the same form. This hints that the above
infrared logarithms may be organized into the Sudakov exponents
with the different upper bounds $Q$ and $m_c$; namely, the Sudakov
exponent $s_c(Q,b)$ for an energetic charm quark up to next-to-leading-logarithm
might be expressed as the difference
\begin{eqnarray}
s_c(Q,b)&=&s(Q,b)-s(m_c,b),\nonumber\\
&=&\int_{m_c}^Q\frac{d \mu}{\mu}
\left[\int_{1/b}^{\mu}\frac{d\bar\mu}{\bar\mu}A(\alpha_s(\bar\mu))
+B(\alpha_s(\mu))\right].\label{sc}
\end{eqnarray}
This observation applies to the organization of the double logarithms in
Figs.~\ref{fig:fig3}(b) and \ref{fig:fig3}(c).

At last, the $\mu_f$-dependent logarithm $\ln(\mu_f^2/m_c^2)$ in Eq.~(\ref{mass})
means that the $J/\psi$ (as well as $B_c$) meson distribution amplitude
is defined at the scale $m_c$ and that the
renormalization-group evolution for the $B_c\to J/\psi$ transition runs from $\mu_f=m_c$
to the hard scale of the process. We summarize the exponents of the total evolution
factors for the $B_c$ and $J/\psi$ meson distribution amplitudes as
\begin{eqnarray}
S_{B_c}&=&s_c\left(x_1P_1^-,b_1\right)+\frac{5}{3}\int^t_{m_c}\frac{d\bar\mu}{\bar\mu}
\gamma_q(\alpha_s(\bar\mu)),\nonumber\\
S_{J/\psi}&=&s_c\left(x_2P_2^+,b_2\right)+s_c\left((1-x_2)P_2^+,b_2\right)+
2\int^t_{m_c}\frac{d\bar\mu}{\bar\mu}
\gamma_q(\alpha_s(\bar\mu)),\label{sbc}
\end{eqnarray}
with the hard scale $t$, and the quark anomalous dimension $\gamma_q=-\alpha_s/\pi$,
that governs the aforementioned renormalization-group evolution. The coefficient $5/3$
in the first line of Eq.~(\ref{sbc}) differs from the coefficient 2 in the second
line, since we have employed the effective heavy quark field for the bottom quark
in the definition of the $B_c$ meson distribution amplitude, as exhibited by the
horizontal double line in Fig.~\ref{fig:fig3}(c).
For the numerical analysis below, we insert the one-loop running coupling constant
$\alpha_s$ into Eq.~(\ref{sbc}) in order to match the expected
next-to-leading-logarithm accuracy of our resummation formula.

\section{\boldmath $B_c^+ \to J/\psi \pi^+$ decay}

\begin{figure}[!!htb]
\centering
\begin{tabular}{l}
\includegraphics[width=0.8\textwidth]{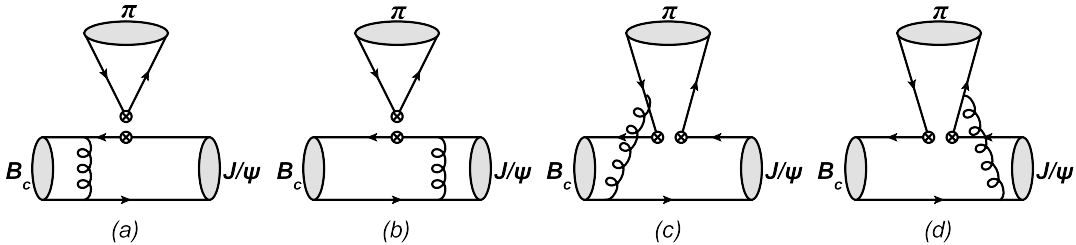}
\end{tabular}
\caption{Leading-order diagrams for the $B_c^+ \to J/\psi \pi^+$ decay in the PQCD approach.}
\label{fig:fig4}
\end{figure}

After the pioneering paper on $B_c$ meson decays by Bjorken in
1986~\cite{Bjorken:1986kfa}, numerous investigations in different formalisms
have been devoted to this subject, but the predictions
vary in a wide range. For example, the $B_c^+ \to J/\psi \pi^+$ branching
ratio was predicted to be between orders of $10^{-4}$ and
$10^{-2}$~\cite{Bc2PsiPi,Sun:2007ei,Sun:2008ew,Rui:2014tpa,Rui:2016opu}.
In particular, it takes the values $1.2 \times 10^{-3}$ in the QCD factorization
approach~\cite{Sun:2007ei}
and $(1.4 \sim 2.5) \times 10^{-3}$~\cite{Sun:2008ew},
$2.33^{+0.63+0.16+0.48}_{-0.58-0.16-0.12} \times 10^{-3}$~\cite{Rui:2014tpa}, and
$2.6^{+0.6+0.2+0.8}_{-0.4-0.2-0.2} \times 10^{-3}$~\cite{Rui:2016opu} in the
conventional PQCD approach. These results manifest the
sensitivity to the hadronic inputs in the theoretical frameworks for $B_c$
meson decays. However, the current data, appearing only as the ratios of the
decay rates because of experimentally complicated background, such as
\begin{eqnarray}
R_{K/\pi}^{J/\psi} \equiv \frac{Br(B_c \to J/\psi K^+)}{Br(B_c \to J/\psi \pi^+)},
\end{eqnarray}
cannot be used to discriminate the branching-ratio predictions.
The factorizable emission diagrams in Figs.~\ref{fig:fig4}(a)
and \ref{fig:fig4}(b) dominate the $B_c^+ \to J/\psi K^+$ and
$B_c^+ \to J/\psi \pi^+$ modes, so the associated uncertain
$B_c \to J/\psi$ transition form factor cancels in the ratio.
This explains why the various formalisms lead to similar
$R_{K/\pi}^{J/\psi}$ in agreement with the latest measurement~\cite{Aaij:2016tcz},
although they give quite distinct values for the individual branching
ratios.

In this section, we calculate the $B_c \to J/\psi$ transition form factor and
the $B_c^+ \to J/\psi \pi^+$ branching ratio in the improved PQCD approach
developed in Sec.~III. The relevant weak effective Hamiltonian $H_{{\rm eff}}$
is written as~\cite{Buchalla:1995vs}
\beq
H_{\rm eff}\, &=&\, {G_F\over\sqrt{2}}
V^*_{cb}V_{ud} [ C_1(\mu)O_1(\mu)
+C_2(\mu)O_2(\mu) ]+ {\rm H.c.}\;,
\label{eq:heff}
\eeq
where $C_{1,2}(\mu)$ are the Wilson coefficients evaluated at the renormalization
scale $\mu$ and the local four-quark operators are
\beq
O_1\, &=&\,
\bar{d}_\alpha \gamma_\mu (1- \gamma_5) u_\beta \;\
\bar{c}_\beta \gamma_\mu (1- \gamma_5) b_\alpha,
\qquad
O_2\, =\, \bar{d}_\alpha \gamma_\mu (1- \gamma_5) u_\alpha\;\
\bar{c}_\beta \gamma_\mu (1- \gamma_5) b_\beta,
\label{eq:operators-1}
\eeq
with the color indices $\alpha$ and $\beta$ and the Fermi constant
$G_F=1.16639\times 10^{-5}$ ${\rm GeV}^{-2}$.
For the Cabibbo-Kobayashi-Maskawa matrix elements $V_{cb}$ and $V_{ud}$,
we employ the Wolfenstein parametrization at leading order with the
parameters $A =0.811$ and $\lambda =0.22506$~\cite{Olive:2016xmw}.
The momenta of the $B_c$ and $J/\psi$ mesons have been chosen in
Eq.~(\ref{bj}), from which the pion momentum is given by
$P_3 = m_{B_c}/\sqrt{2} (0, 1- r_{J/\psi}^2, {\bf 0}_T)$,
for the vanishing pion mass.
The momenta of the spectator quarks in the involved hadrons are parametrized as
\begin{eqnarray}
k_1 = (x_1 P_1^+, x_1 P_1^-, {\bf k}_{1T}) , \qquad
k_2 = (x_2 P_2^+, x_2 P_2^-, {\bf k}_{2T}), \qquad
k_3= (x_3 P_3^+, x_3 P_3^-, {\bf k}_{3T}).
\end{eqnarray}

The $B_c$, $J/\psi$, and $\pi$ meson distribution amplitudes have the structures
\beq
\Phi_{B_c}(x,b)&\equiv& \frac{i}{\sqrt{2 N_c}}(\psl_{B_c}  +m_{B_c})\gamma_5
 \phi_{B_c}(x,b) ,
\label{eq:def-bc-wf} \\
\Phi_\pi(x) &\equiv& \frac{i}{\sqrt{2 N_c}} \gamma_5
\left[\psl_\pi \;\phi_\pi^A(x)+ m_0^\pi \phi_\pi^P(x) + m_0^\pi (\nsl\, \vsl -1)
\phi_\pi^T(x)\right],
\label{eq:def-pion-wf}\\
\Phi^{L}_{J/\psi}(x) &\equiv&  \frac{1}{\sqrt{2 N_c}}
   \left[ m_{J/\psi}\, {\epsl}_{J/\psi}^L \,\phi_{J/\psi}^L(x)  +
 {\epsl}_{J/\psi}^L \, \psl_{J/\psi} \;\phi_{J/\psi}^t(x)  \right],
 \label{eq:def-jpsi-wf}
\eeq
with the dimensionless vectors $n=(0,1,{\bf 0}_T)$ and $v=(1,0,{\bf 0}_T)$
and the longitudinal polarization vector for the $J/\psi$ meson
\begin{eqnarray}
\epsilon_{J/\psi}^L = \frac{1}{\sqrt{2} r_{J/\psi}}(1, -r_{J/\psi}^2, {\bf 0}_T).
\end{eqnarray}
Owing to the experimental status stated before, we adopt the
shape parameter $\beta_{B_c}=1$ GeV for the $B_c$ meson distribution
amplitude inferred from the kinematic constraint.
The light-cone pion distribution amplitudes
$\phi_{\pi}^A$ (twist 2), and $\phi_{\pi}^P$ and
$\phi_{\pi}^T$ (twist 3) have been parametrized
as~\cite{Chernyak:1983ej,Ball:1998tj,Braun:2004vf}
\beq
\phi_{\pi}^A(x) &=& \frac{f_{\pi}}{2\sqrt{2N_c}}\, 6x(1-x) \left[1
+ a_2^{\pi}C_2^{3/2}(2x-1)+a_4^{\pi}C_4^{3/2}(2x-1)\right] ,
\label{eq:pionda-A}
\\
\phi^P_{\pi}(x) &=& \frac{f_{\pi}}{2\sqrt{2N_c}}\, \bigg[ 1
+\left(30\eta_3 -\frac{5}{2}\rho_{\pi}^2\right) C_2^{1/2}(2x-1) \non &
& \hspace{35mm} -\, 3\left( \eta_3\omega_3 +
\frac{9}{20}\rho_{\pi}^2(1+6a_2^{\pi}) \right) C_4^{1/2}(2x-1)
\bigg],
\label{eq:pionda-P}\\
\phi^T_{\pi}(x) &=& \frac{f_{\pi}}{2\sqrt{2N_c}}\,
(1-2x)\bigg[ 1 + 6\left(5\eta_3 -\frac{1}{2}\eta_3\omega_3 -
\frac{7}{20}
      \rho_{\pi}^2 - \frac{3}{5}\rho_{\pi}^2 a_2^{\pi} \right)
(1-10x+10x^2) \bigg],
\label{eq:pionda-T}
\eeq
with the decay constant $f_\pi = 0.130$ GeV; the Gegenbauer moments
$a_2^{\pi}= 0.115 \pm 0.115$ and
$a_4^{\pi}=-0.015$; the parameters $\eta_3=0.015$ and
$\omega_3=-3$~\cite{Chernyak:1983ej,Ball:1998tj}; the mass ratio
$\rho_{\pi}=m_{\pi}/m_{0}^{\pi}$, $m_0^\pi=1.4$ GeV being the pion
chiral mass; and the Gegenbauer polynomials $C_n^{\nu}(t)$,
\begin{eqnarray}
C_2^{1/2}(t)\,& =&\, \frac{1}{2} \left(3\, t^2-1\right) ,\;\; \qquad
C_4^{1/2}(t)\, =\, \frac{1}{8} \left(3-30\, t^2+35\, t^4\right) ,
\nonumber\\
C_2^{3/2}(t)\, &=&\,
\frac{3}{2} \left(5\, t^2-1\right) , \;\; \qquad C_4^{3/2}(t) \,=\,
\frac{15}{8} \left(1-14\, t^2+21\, t^4\right) .
\end{eqnarray}
The $J/\psi$ meson distribution amplitudes $\phi_{J/\psi}^L$ (twist 2) and
$\phi_{J/\psi}^t$ (twist 3) have been derived as~\cite{Bondar:2004sv}
\begin{eqnarray}
\phi_{J/\psi}^L(x)&=& 9.58\frac{f_{J/\psi}}{2\sqrt{2N_c}}x(1-x)
\left[\frac{x(1-x)}{1-2.8x(1-x)}\right]^{0.7},
\label{eq:psida-L}\\
\phi_{J/\psi}^t(x)&=&10.94\frac{f_{J/\psi}}{2\sqrt{2N_c}}(1-2x)^2
\left[\frac{x(1-x)}{1-2.8x(1-x)}\right]^{0.7},
\label{eq:psida-t}
\end{eqnarray}
with the decay constant $f_{J/\psi}=0.405 \pm 0.014$ GeV.

The $B_c^+ \to J/\psi \pi^+$ decay amplitude is decomposed into
\beq
{\cal A}(B_c \to J/\psi \pi) &=& V_{cb}^* V_{ud} (f_\pi F+  M ).
\eeq
The factorizable emission diagrams, i.e., Figs.~\ref{fig:fig4}(a)
and \ref{fig:fig4}(b), give the factorization formula
\beq
F &=& 8 \pi C_F m_{B_c}^2 \int_0^1 dx_1 dx_2
\int_0^\infty b_1db_1 b_2db_2 \phi_{B_c}(x_1,b_1) (r_{J/\psi}^2 -1)
\non && \times
\biggl\{  \left[ r_{J/\psi}(r_b +2 x_2 -2) \phi_{J/\psi}^t(x_2)
-
(2 r_b + x_2 -1) \phi_{J/\psi}^L(x_2)  \right] h_a(x_1,x_2,b_1,b_2) E_f(t_a)
 \non &&
+
\left[r^2_{J/\psi}(x_1 -1)-r_c \right]\phi_{J/\psi}^L(x_2) h_b(x_1,x_2,b_1,b_2)
E_f(t_b)  \biggr\} ,
\eeq
where the ratios $r_b= m_b/m_{B_c}$ and $r_c = m_c/ m_{B_c}$ and $b_i$ are
the impact parameters conjugate to the transverse momenta
$k_{iT}$. It is known that the above formula
is related to the transition form factor
$A_0^{B_c \to J/\psi}(q^2=0)$~\cite{Keum:2000wi,Lu:2000em,Lu:2000hj}
with $q=P_{1} -P_{2}$. As pointed out in the Introduction, the PQCD approach
is applicable to the evaluation of the nonfactorizable
emission diagrams, i.e., Fig.~\ref{fig:fig4}(c) and \ref{fig:fig4}(d).
The corresponding factorization formula is expressed as
\beq
M &=& -\frac{32}{\sqrt{6}}\pi C_F m_{B_c}^2 \int_0^1 dx_1 dx_2 dx_3
\int_0^\infty b_1db_1 b_3db_3 \phi_{B_c}(x_1,b_1) \phi_\pi^A(x_3) (r_{J/\psi}^2 -1) \non & & \times
\biggl\{\left[(r_{J/\psi}^2-1)(x_1+x_3-1)\phi_{J/\psi}^L(x_2)
+r_{J/\psi}(x_2 -x_1)\phi_{J/\psi}^t(x_2)\right] h_c(x_1,x_2,x_3,b_1,b_3) E_f(t_c)
\non &&
+\left[(2x_1-(x_2+x_3)+r_{J/\psi}^2(x_3-x_2))\phi_{J/\psi}^L(x_2) + r_{J/\psi}(x_2-x_1)\phi_{J/\psi}^t(x_2)\right] h_d(x_1,x_2,x_3,b_1,b_3)E_f(t_d)\biggr\}.
 \eeq

In the above expressions, the hard functions $h_{a,b,c,d}$ are defined by
\beq
h_a(x_1,x_2,b_1,b_2) &=& \left[
\theta(b_2-b_1)I_0(\sqrt{\beta_a}b_1) K_0(\sqrt{\beta_a}b_2)
+(b_1 \leftrightarrow b_2)\right]K_0(\sqrt{\alpha}b_1), \\
h_b(x_1,x_2,b_1,b_2) &=& \left[
\theta(b_2-b_1)I_0(\sqrt{\beta_b}b_1) K_0(\sqrt{\beta_b}b_2)
+ (b_1 \leftrightarrow b_2)\right]K_0(\sqrt{\alpha}b_2),\\
 h_{c,d}(x_1,x_2,x_3,b_1,b_3) &=&
 \left[\theta(b_3-b_1) I_0(\sqrt{\alpha} b_1)
 K_0(\sqrt{\alpha} b_3)
+(b_1 \leftrightarrow b_3) \right]
K_0( \sqrt{\beta_{c,d}} b_3),
\label{eq:pp1}
 \eeq
with the factors $\alpha$ and $\beta_{a,b,c,d}$ and the
hard scales $t_{a,b,c,d}$,
\beq
\alpha &=& -[(x_1-x_2)(x_1-x_2 r_{J/\psi}^2)]m_{B_c}^2,\\
\beta_a &=& -[(1-x_2)(1-x_2 r_{J/\psi}^2)-r_b^2]m_{B_c}^2,\qquad
\beta_b = -[(1-x_1)(r_{J/\psi}^2-x_1)-r_c^2]m_{B_c}^2,\\
\beta_c &=& -[(x_2 r_{J/\psi}^2+(1-x_3)(1-r_{J/\psi}^2)-x_1)
(x_2-x_1)]m_{B_c}^2 , \\
\beta_d &=& -[(x_2 r_{J/\psi}^2+x_3 (1-r_{J/\psi}^2)-x_1)
(x_2-x_1)]m_{B_c}^2,\\
t_a &=& \max(\sqrt{|\alpha|},\sqrt{|\beta_a|},1/b_1,1/b_2),\qquad
t_b = \max(\sqrt{|\alpha|},\sqrt{|\beta_b|},1/b_1,1/b_2),\\
t_c &=& \max(\sqrt{|\alpha|}, \sqrt{|\beta_c|}, 1/b_1, 1/b_3) , \qquad
t_d = \max(\sqrt{|\alpha|}, \sqrt{|\beta_d|}, 1/b_1, 1/b_3).
\eeq
Note that, as $\alpha$ and $\beta_{a,b,c,d}$ are negative, the associated
Bessel functions transform as
\begin{eqnarray}
K_0(\sqrt{y}) = K_0(i\sqrt{|y|})= \frac{i \pi}{2} [J_0(\sqrt{|y|})
+ i N_0(\sqrt{|y|})] \;, \qquad I_0(\sqrt{y}) = J_0(\sqrt{|y|}),  \hspace{0.5cm}
\end{eqnarray}
for $y<0$. The evolution functions $E_f(t)=\alpha_s(t)C_i(t) S_i(t) $ contain
the Wilson coefficients
\beq
  C_{ab}(t) &=&  \frac{1}{3} C_1(t) + C_2(t) ,  \qquad
  C_{cd}(t) = C_1(t)
\eeq
and the Sudakov factors
\beq
S_{ab}(t)
&=& s_c\left(x_1 P_1^-, b_1\right) +s_c\left(x_2
P_2^+, b_2\right) +s_c\left((1-x_2) P_2^+,
b_2\right) -\frac{1}{\beta_1}\left[\frac{11}{6} \ln\frac{\ln(t/\Lambda)}{\ln(m_c/\Lambda)}
\right],
\label{eq:sab}\\
S_{cd}(t) &=& s_c\left(x_1 P_1^-, b_1\right)
 +s_c\left(x_2 P_2^+, b_1\right)
+s_c\left((1-x_2) P_2^+, b_1\right) +s\left(x_3
P_3^-, b_3\right) +s\left((1-x_3) P_3^-,
b_3\right) \non
 && -\frac{1}{\beta_1}\left[
\frac{11}{6} \ln\frac{\ln(t/\Lambda)}{\ln(m_c/\Lambda)}
+\ln\frac{\ln(t/\Lambda)}{-\ln(b_3\Lambda)}\right]
\label{eq:scd}
\eeq
where the explicit expression of the Sudakov exponent $s(Q,b)$
for an energetic light quark is
referred to Refs.~\cite{Keum:2000wi,Lu:2000em}.

With the QCD scale $\Lambda_{\rm QCD}^{(4)} = 0.25$ GeV and the $B_c$ meson
lifetime $\tau_{B_c}=0.507$ ps, we obtain
$Br(B_c^+ \to J/\psi \pi^+)=1.60 \times 10^{-3}$. This result is consistent with
$1.2 \times 10^{-3}$ derived in the QCD factorization approach~\cite{Sun:2007ei},
in which the transition form factor $A_0^{B_c \to J/\psi}$ was treated as an input,
a bit larger value of $A_0^{B_c \to J/\psi}=0.6$ was employed,
and the one-loop correction to the $b\to c$ decay vertex was included.
Our prediction can be compared to
the measured branching ratio of the corresponding mode with the replacement
of the spectator charm quark by an up quark,
$Br(B^+ \to \bar{D}^{*0} \pi^+)=(5.18 \pm 0.26) \times 10^{-3}$~\cite{Olive:2016xmw},
which receives an additional color-suppressed tree contribution.
The dependence of the
quantities $A_0^{B_c \to J/\psi}(0)$ and $Br(B_c^+ \to J/\psi \pi^+)$ on
$\beta_{B_c}$ in the range $[0.8, 1.2]$ GeV is shown in
Table~\ref{tab:d-beta-A0-Br}. It is clearly seen that the imaginary piece
of the $B_c \to J/\psi$ transition form factor is greatly suppressed, being
only $10\%-20\%$ of the real piece, and that the $B_c^+ \to J/\psi \pi^+$ branching ratio
is unlikely to be lower than $10^{-3}$. Roughly speaking, the preferred range of
$Br(B_c^+ \to J/\psi \pi^+)$ from the PQCD approach
can be preliminarily read as [0.9, 2.8]$\times 10^{-3}$. When the data are available
for individual branching ratios, or for the ratios of decay rates that are
more sensitive to the nonfactorizable emission contributions, it is possible
to pin down the shape parameter $\beta_{B_c}$ and to make more precise
predictions in the PQCD approach. In the latter case, the emitted meson could
be a scalar or tensor, such that the dominant nonfactorizable emission
diagrams do not cancel in the ratios of decay rates.

\begin{table}[htb]
\caption{Dependence on the shape parameter $\beta_{B_c}$
of the quantities $A_0^{B_c \to J/\psi}(0)$ and
$Br(B_c^+ \to J/\psi \pi^+)$ in the improved PQCD formalism. }
\label{tab:d-beta-A0-Br}
 \begin{center}\vspace{-0.5cm}{
\begin{tabular}[t]{c||c|c}
\hline  \hline
   shape parameter   &  $A_0^{B_c \to J/\psi}(0)$  & $Br(B_c^+ \to J/\psi \pi^+)$ \\
\hline
 $\beta_{B_c} =0.8$~ GeV
     &$0.488 - {\it i} 0.095$
     &$2.80 \times 10^{-3}$
 \\
 $\beta_{B_c} = 0.9$~ GeV
     &$0.434 - {\it i} 0.070$
     &$2.10 \times 10^{-3}$
 \\
 $\beta_{B_c} = 1.0$~ GeV
     &$0.384 - {\it i} 0.053$
     &$1.60 \times 10^{-3}$
 \\
 $\beta_{B_c} = 1.1$~ GeV
     &$0.341 - {\it i} 0.039$
     &$1.23 \times 10^{-3}$
 \\
 $\beta_{B_c} = 1.2$~ GeV
     &$0.306 - {\it i} 0.029$
     &$0.94 \times 10^{-3}$
 \\
 \hline \hline
\end{tabular}}
\end{center}
\end{table}

\bigskip

\section{CONCLUSION}

In this paper, we have deduced the shape of the $B_c$ meson distribution
amplitude $\phi_{B_c}(x)$ resulting from the soft gluon emission effect based
on the parton kinematic analysis and found that $\phi_{B_c}(x)$ exhibits a peak
around the momentum fraction $x\sim m_c/m_b\sim 0.3$ of the spectator charm
quark with a width of order $\Lambda/m_b\sim 0.1$. These features were then
implemented into the parametrization of $\phi_{B_c}(x)$ in terms of a Gaussian
form with the shape parameter $\beta_{B_c}\sim 1.0$ GeV. We have estimated the
potential imaginary piece in the $B_c\to J/\psi$ transition form factor, which
should be power suppressed according to the specified parton kinematics
and the argument on the absence of intermediate states.
It is worth emphasizing that the resummation formula adopted in the
conventional PQCD approach to $B_c$ meson decays~\cite{Sun:2008ew,Rui:2014tpa,Rui:2016opu}
is not appropriate. We have modified the $k_T$ resummation by taking into account
the finite charm quark mass, the effect of which was shown to enhance
the decay rates. We point out that this modification is exact only at the
leading-logarithm level, and a precise next-to-leading-logarithm
resummation formalism for a hadronic process
involving the multiple scales $m_b$, $m_c$, and $\Lambda_{\rm QCD}$ is still urged;
it demands a complete one-loop calculation for determining the factor
$B(\alpha_s)$ in Eq.~(\ref{light}).

Given the $B_c$ meson distribution amplitude preferred by the kinematic
constraints and the newly derived Sudakov factor for the $B_c\to J/\psi$
transition, we have calculated, at leading order in the strong coupling,
the transition form factor $A_0^{B_c \to J/\psi}(0)$ and the
$B_c^+ \to J/\psi \pi^+$ branching ratio in the range [0.8,1.2] GeV of the
shape parameter $\beta_{B_c}$. It was observed that the strong phase in
$A_0^{B_c \to J/\psi}(0)$ is indeed largely suppressed and that
the predicted $Br(B_c^+\to J/\psi\pi^+)\sim 1.60\times 10^{-3}$ is
comparable to the data $Br(B^+ \to\bar{D}^{*0} \pi^+)=(5.18 \pm 0.26) \times 10^{-3}$.
The definite value of the shape parameter demands the input data of some individual
$B_c$ decay channels from LHCb, with which it is then possible to make more
precise predictions for various modes.
At last, we stress that the improved PQCD formalism developed in this work
is applicable to $B_c$ meson decays to other charmonia and charmed mesons.

\begin{acknowledgments}

We thank Y.L.~Shen and R.L.~Zhu for valuable discussions.
This work is supported in part by
the Ministry of Science and Technology of R.O.C. under
Grant No. MOST-104-2112-M-001-037-MY3, by the National Natural Science
Foundation of China under Grants No.~11765012 and No.~11775117,
and by the Research Fund of
Jiangsu Normal University under Grant No.~HB2016004.
\end{acknowledgments}


\end{CJK*}
\end{document}